# Low Temperature-Mediated Enhancement of Photoacoustic Imaging Depth

Sadreddin Mahmoodkalayeh[1,2,5¥], Hossein Z. Jooya[3¥], Ali Hariri[1¥], Yang Zhou[1], Qiuyun Xu[1], Mohammad A. Ansari[5], and Mohammad R.N. Avanaki[1,4,6,*]

[1] *Department of Biomedical Engineering, Wayne State University, Detroit, Michigan, USA*

[2] *Department of Physics, Shahid Beheshti University, Tehran, Iran*

[3] *Harvard-Smithsonian Center for Astrophysics, Harvard University, Cambridge, MA*

[4] *Department of Dermatology, Wayne State University School of Medicine, Detroit, Michigan, USA*

[5] *Laser and Plasma Research Institute, Shahid Beheshti University, Tehran, Iran*

[6] *Barbara Ann Karmanos Cancer Institute, Detroit, Michigan*

[*]*Corresponding author: mrn.avanaki@wayne.edu*

[¥]*These authors have contributed equally in this work.*

**Abstract**

We study the temperature dependence of the underlying mechanisms related to the signal strength and imaging depth in photoacoustic imaging. The presented theoretical and experimental results indicate that imaging depth can be improved by lowering the temperature of the intermediate medium that the laser passes through to reach the imaging target. We discuss the temperature dependency of optical and acoustic properties of the intermediate medium and their changes due to cooling. We demonstrate that the SNR improvement of the photoacoustic signal is mainly due to the reduction of Grüneisen parameter of the intermediate medium which leads to a lower level of background noise. These findings may open new possibilities toward the application of biomedical laser refrigeration.

**Introduction**

Photoacoustic Imaging (PAI) is a promising functional and molecular imaging technique for a wide range of biomedical applications [1]. PAI combines the technological advances of both optical and acoustic imaging, i.e., the high intrinsic contrast of optical imaging and the spatial resolution of ultrasound imaging [2, 3, 49]. Every material, including bodily substances, has a specific optical absorption coefficient unique to endogenous chromospheres in cells [2, 3]. The substance to be imaged is illuminated by a nanosecond pulsed laser of a specific wavelength which correlates to the highest absorption coefficient of the subject [4-7]. The amplitude of a photoacoustic signal, $A_{PA}$, is proportional to the light fluence, $F$, an acoustic attenuation coefficient, $\xi$, the optical absorption coefficient of the imaging target, $\mu_a$, and, a temperature dependent coefficient, $\Gamma$, known as Grüneisen parameter (Eq.1).

$$A_{PA} = \xi \Gamma \mu_a F \tag{1}$$

The temperature dependency of the Grüneisen parameter has previously been explored in several studies [8, 10-18, 20]. There are also studies showing the temperature dependency of the scattering coefficient of the tissue [19-24]. There is not sufficient evidence about how the absorption coefficient of tissue alters with temperature, particularly in the range of temperature investigated in our study [22-24].

Recently several research groups have developed temperature-dependent PA systems based on the dependence of Grüneisen parameter to temperature, in biological tissues [25-28]. Following thermal nonlinearity theorem, Simandoux *et al.*, investigated the thermal-based nonlinear PA generation to discriminate between different types of absorbers [29]. Zharov developed an ultrasharp nonlinear photothermal (PT) and PA spectroscopy system, which enables an enhancement in specificity and sensitivity of PT/PA spectral analysis [30]. Based on the same principle, Tian *et al.* developed a dual-pulse nonlinear photoacoustic technique [31]. Wang *et al.*, developed a Grüneisen relaxation photoacoustic microscopy (GR-PAM) system in which two laser pulses with a specified time delay were employed, one for thermal tagging, the other one for signal generation [8]. They showed that when the second laser pulse excites the tagged absorbers within the thermal relaxation time, a stronger photoacoustic signal than the initial one is obtained. In another configuration, they used a continuous wave (CW) laser for thermal tagging and observed that the performance of their method is diminished. Although this scheme enhances the photoacoustic signal generated from the imaging target, the same mechanism can compromise the penetration depth due to the undesired thermal excitation of intermediate medium.

There have been studies that have utilized the temperature dependence of Grüneisen parameter to improve the PA image by increasing the temperature of the imaging target without increasing the temperature of the intermediate medium [8]. Here, we propose to decrease the temperature of the intermediate medium while keeping the temperature of the imaging target constant. In biological tissues due to multiple scattering, the fluence is significantly affected and the photons are greatly scattered especially after they reach the diffusion limit (i.e., ~ 1mm). Such scattered photons generate background PA signals which reduce the signal-to-noise ratio (SNR) of the PA signal generated from the imaging target. Cooling the intermediate medium reduces such background signals, and increases the SNR of the PA signal of the imaging target.

In this study, we report the theoretical considerations and experimental observations for the SNR enhancement in photoacoustic imaging by cooling the intermediate medium. The proposed cooling mechanism opens up the possibility of using the laser refrigeration technology in the future. This approach creates an efficient channel through the biological tissue layers, which in turn, can significantly improve the penetration depth in photoacoustic imaging.

**Theoretical background**

The process of signal acquisition in photoacoustic imaging is comprised of three steps: laser light illumination, conversion of deposited optical energy to initial pressure via thermoelastic expansion, and a broad-band PA wave propagation through the medium. In PA imaging, the initial pressure, $p_0$, is generated according to Eq.2. In Eq.2, the Grüneisen parameter (dimensionless) relates the absorbed energy to the changes in the pressure.

$$p_0(r) = \Gamma(r)\mu_a(r)F(r) \qquad (2)$$

Biological tissues are highly scattering in visible and near infrared (NIR) range. Light propagation has a sophisticated behavior in tissues, and can be described by the radiative transfer equation [32] and its approximation, i.e., the diffusion equation [33, 34]. Different tissues are characterized by their optical properties, namely, absorption coefficient, scattering coefficient, anisotropy, and refractive index. As light passes through the tissue, it is attenuated due to the absorption and scattering. Hence, any change in these parameters can alter the fluence; leading to a change in the PA signal amplitude.

Grüneisen parameter is defined as:

$$\Gamma = V\left(\frac{dP}{dE}\right)_V = \frac{\beta v_s^2}{C_p} \qquad (3)$$

Where $\beta$ is the thermal expansion coefficient, $v_s$ is the speed of sound in the medium and $C_p$ is the heat capacity at a constant pressure. There is also an acoustic attenuation coefficient, $\xi$, which accounts for the PA wave attenuation as it travels through the medium. Temperature change can alter the optical and acoustic properties of the tissue, and thus the PA signal strength.

Effect of cooling of the intermediate medium on the SNR improvement can be explained using the method proposed by Steven L. Jacques [55]. The background noise level is determined by calculating the velocity potential at different temperatures. We consider a tissue with a uniform speed of sound, $v_s$, inside which there is an imaging target. There is an ultrasound transducer placed on the surface of the tissue. PA waves generated from the target at the distance $r$ from the transducer reach the transducer at time $t=r/v_s$. The PA waves from a hemispherical shell with a radius $r$ and a thickness $dr = v_s dt$ reach the transducer at times between $t - dt/2$ to $t + dt/2$. The noise, $PA_{noise}$, is calculated from the PA contribution of the shell that is not in the imaging region, as described by equation (4).

$$PA_{noise} = -\rho \, d\phi_{background}/dt \tag{4}$$

where $\rho$ represents the density and $\phi_{background}$ the velocity potential.

$$-\phi_{background}(t) = \frac{\beta}{4\pi\rho c_p} \frac{1}{dt} \int_{r-dr/2}^{r+dr/2} \frac{E_{shell,background}}{r} 4\pi r^2 dr = \frac{\Gamma}{4\pi\rho v_s^2} \frac{1}{dt} \int_{r-dr/2}^{r+dr/2} \frac{E_{shell,background}}{r} 4\pi r^2 dr \tag{5}$$

where $\beta$ is the thermal expansion coefficient, $E_{shell,background}$ is the averaged deposited light energy on the shell outside the target region. The $PA_{noise}$ that reaches the transducer at time $t$ is:

$$PA_{noise}(t) = \Gamma_{outside} E_{shell,background} \tag{6}$$

$PA_{signal}$ is calculated from the PA contribution of points on the shell that are in the imaging region and can be calculated as:

$$PA_{signal}(t) = \Gamma_{inside} E_{shell,target} \tag{7}$$

where $E_{shell,target}$ is the averaged deposited light energy on the shell inside the target region.

If the optical parameters, absorption and scattering coefficients, are temperature independent, the deposited energy will be the same at different temperatures and the PA signal generated will only depend on the Grüneisen parameter. If the temperature of the target region is constant, the Grüneisen parameter and consequently the $PA_{signal}$ will be constant. Decreasing the temperature of the intermediate medium will decrease the Grüneisen parameter and consequently the $PA_{noise}$. SNR is defined as equation (8),

$$SNR = 10\log\left|\frac{PA_{signal}}{PA_{noise}}\right| \tag{8}$$

The changes in the SNR due to temperature change ($\Delta T = T_2 - T_1$) is:

$$\Delta SNR = SNR(T_2) - SNR(T_1) = 10\log\left|\frac{PA_{noise}(T_1)}{PA_{noise}(T_2)}\right| = 10\log\left|\frac{\Gamma_{outside}(T_1)}{\Gamma_{outside}(T_2)}\right| \qquad (9)$$

It can be seen that if the temperature of the intermediate medium - the region outside the target – decreases, the SNR of the PA signal belong to the imaging target will be improved.

The results obtained from an ultrasound imaging study done by Gerter *et al*., [35] and Clarke *et al.*, [36] on the liver tissue showed no significant change in acoustic attenuation coefficient at temperatures close to 50 ºC; the changes were more significant for higher temperatures. This study did not include the results of lower temperatures. Bamber *et al.*, [37] studied the temperature dependency of the acoustic attenuation coefficient in soft tissues, and showed that increasing the temperature from 10 ºC to 30 ºC, decreases the attenuation coefficient.

There is only a few studies on temperature-dependence of the optical properties. Changes in the reduced scattering coefficient in temperatures above 55ºC are mostly related to substantial structural changes in the tissue, e.g., proteins coagulation [14, 38, 39], which is likely irreversible. Such changes do not occur at lower temperatures (< 55ºC). Laufer *et al.* reported a 0.5% ºC$^{-1}$ reduced scattering coefficient increase in dermis and 0.14% ºC$^{-1}$ decrease in sub-dermis with changing the temperature from 25 to 40ºC [24]. They related different temperature dependency schemes to compositions of skin layers; dermis is largely composed of protein, and sub-dermis fat. Cletus *et al.* found a negative scattering coefficient change with temperature rise for intralipid [21], and Jaywant *et al.* reported a positive scattering coefficient change for bovine muscle [23]. They both are consistent with Laufer's findings. Jaywant also found that there is only a slight change in the scattering coefficient of bovine brain tissue with the temperature rise [23] which can be due to the opposite effects of fat and protein scattering coefficient change with the temperature rise. Troy *et al.* reported a positive scattering coefficient change for prostate tissue [22]. Yeo *et al.* studied the porcine and found an increase in the scattering coefficient with the temperature rise [20]. Ouyang *et al.* studied the rat skin tissue and found a decrease in the scattering coefficient with temperature rise [8]. Jaywant [23] *et al.* observed a positive scattering coefficient change with temperature rise in the liver tissue, while Larina *et al.* [17] reported no change in the optical attenuation of liver tissue. Although, most of the studies reported changes in the scattering coefficient with the temperature change, are in agreement with each other, there are some discrepancies in the literature. The reason for the discrepancies could be due to different condition of the tissue samples or the wavelength of light they have used.

We use chicken breast tissue in our experiments as an intermediate medium, which contains protein. According to the literature [23, 24], for such tissue, we expect a decrease in the scattering coefficient as the temperature drops. With respect to absorption coefficient, there was no study showing a significant change in absorption coefficient of the tissue with temperature change [8, 17, 22-24]. There is evidence that the absorption coefficient of water is changed with temperature [40, 41]. However, water is not the main absorber in the tissue and the absorption coefficient of tissue cannot be deduced solely based on the temperature dependency of absorption coefficient of water.

There is a large body of literature, claimed that Grüneisen parameter is the major contributor in the PA signal change with the temperature [8, 10-18, 20]. The change of the Grüneisen parameter with the temperature is due to the temperature dependency of the thermal expansion coefficient. Larina *et al.*, showed 1.5% ºC$^{-1}$ increase in the Grüneisen parameter in a biological tissue [17].

**Methods**

Fig. 1 schematically illustrates the experimental setup to use photoacoustic tomography for imaging a target (black covered resistance heating wire) within chicken breast tissue. Quanta-Ray PRO-Series Nd:YAG laser (Spectra-Physics Inc., USA) in addition to VersaScan V1.7 optical parametric oscillator (OPO) (Spectra-Physics Inc., USA), with a pulse width of 7 ns and a repetition rate of 30 Hz is used as the light source. In this experiment, the illumination wavelength of 532 nm has been tested. An optical diffuser (grit number 240, Thorlabs Inc., USA) is used to homogenize the laser beam and make a uniform radiance on the phantom. As indicated in Fig. 1, the laser light is delivered to the tissue using a single large graded index plastic optical fiber (FiberFin Inc., USA) with the diameter of 12 mm and 2 meters length. The orientation of the light illumination is designed to have efficient light delivery to the tissue. To increase the temperature in the imaging target, we connected a DC power supply (PWS4721, Tektronix, USA) directly to it. The imaging target is a black covered resistance heating wire (Nickel-Chromium Alloy, 80% Nickel/ 20% Chromium, Omega®, USA). We control the temperature of the imaging target by adjusting the current on the power supply. A 1 amp current is applied to the imaging target for about 20 seconds so its temperature reaches around 37ºC (body temperature), then decreased to 5 mA. Thermometer 1 (9940N, Taylor

Precision digital panel mount thermometer, USA) is attached to the wire to monitor the temperature of the wire. When the temperature dropped, we applied more current to the imaging target until it reached 37°C. We used an ultrasound linear array, ATL/Philips L7-4 (128 elements), with 5MHz central frequency for PA signal detection. The photoacoustic signals are acquired using 128-channel high frequency ultrasound system (Vantage, Verasonics Inc, USA). The amplitude of the recorded photoacoustic signal from each pulse is first compensated for the pulse-to-pulse fluence fluctuation, then the photoacoustic signal of each pulse is normalized to the maximum photoacoustic signal recorded. We altered the temperature of the water from 5°C to 35°C with 5°C increments. Thermometer 2 (TSP01, Thorlabs, USA) is used in the water tank for monitoring the water temperature. The Q-switch trigger output of the laser is used as a source of trigger to synchronize the photoacoustic data with the temperature acquisition. Since the temperature sensor and PA data acquisition are synchronized, we could readily acquire PA signal associated with a particular temperature. For each temperature increment, we collect ultrasound and photoacoustic images of the imaging target. At each acquisition, we collected 100 frames of US/PA data.

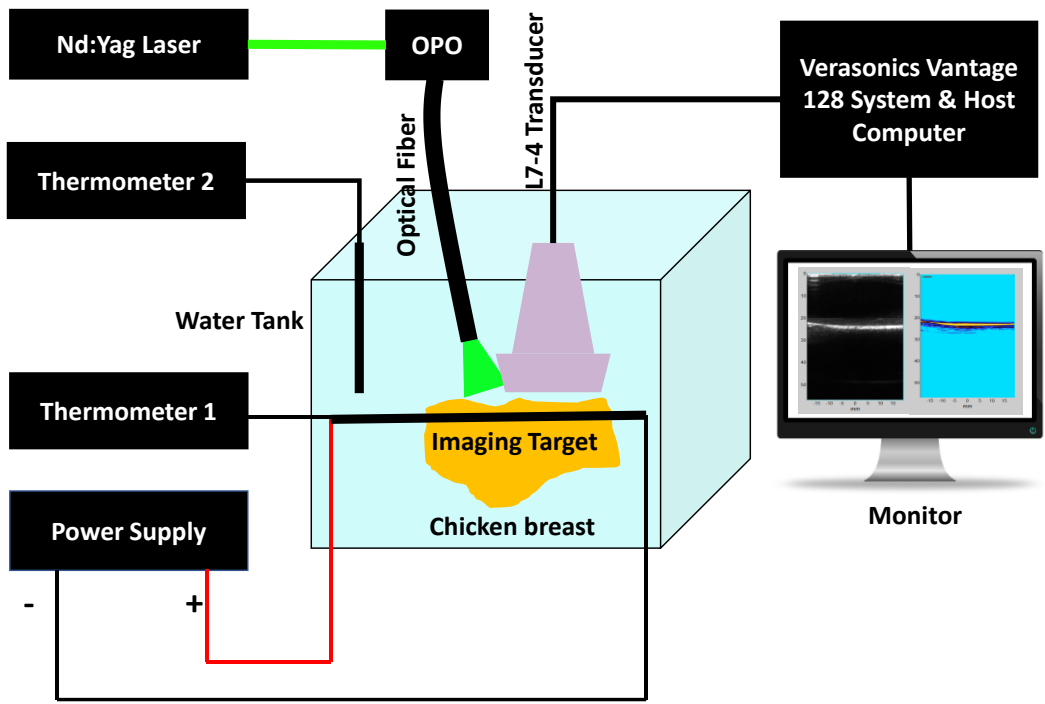

**Figure 1.** Experimental setup of depth-temperature dependency in photoacoustic imaging. The imaging target (black covered resistance heating wire) is placed inside the chicken breast tissue at various distances (10 mm, 20 mm and 30 mm). Photoacoustic images are obtained at temperatures between 5°C and 35°C

with 5°C increments. The imaging target's temperature is controlled and kept constant at 37°C during the experiment.

**Results and Discussion**

**Simulation study**

We performed the simulations at different distances from the illumination source (tissue depths): 5 to 30 mm, at 5 mm increments, and at seven temperatures: 5°C to 35°C, at 5°C increments. Background medium was a slab homogenous tissue, with the optical properties similar to that of chicken breast tissue ($\mu_a = 0.3$ cm$^{-1}$ and $\mu_s' = 10$ cm$^{-1}$) [42]. A rod with 1 mm thickness was placed inside the tissue as the imaging target. Absorption coefficient of the target is set to 30 cm$^{-1}$ to be much greater than intermediate medium. We used a homogenous light with a beam diameter of 1 cm for illumination. A transducer with 4 mm$^2$ area is located on top of the sample and at the center.

The absorbed light energy is calculated by Monte Carlo simulations using MCX software [43]. The initial pressure, $p_0$, is then computed as a product of the absorbed energy and Grüneisen parameter. Photoacoustic waves generated from the target and the intermediate medium are simulated using k-wave software [44].

To calculate the PA signal without any background noise ($PA_{signal}$), we measured the PA signal amplitude generated from the imaging target ($PA_{true}$) embedded in an intermediate medium as well as the PA signal amplitude from the same location when the imaging target is removed, i.e., background signal ($PA_{noise}$). The $PA_{signal}$ is then computed as $PA_{true} - PA_{noise}$. The SNR of the PA signal is calculated by (10):

$$SNR = 10 Log \left|\frac{PA_{signal}(t)}{PA_{noise}(t)}\right|, \quad (10)$$

Where $t=z/v$ is the PA signal arrival time at the transducer plane, $z$ is the depth where object is located at and $v$ is the speed of sound in the intermediate medium. Fig. 2, shows the SNR of the PA signal at different temperatures and depths.

In the first set of simulations, the Grüneisen parameter of the intermediate medium is decreased with the temperature at a rate of $1.5\%°C^{-1}$ [17], while the scattering coefficient of the medium held constant (Fig. 2(a)). The Grüneisen parameter of the target is kept constant since its temperature is not changed. Although the absolute value of SNR is decreasing with depth due to light attenuation, decreasing the temperature of the intermediate medium leads to a SNR improvement. We observed that the PA signal ($PA_{signal}$) does not change with the temperature change of the intermediate medium. This was expected because the temperature (and the Grüneisen parameter) of the imaging target is constant. Moreover, the scattering coefficient of the intermediate medium in this experiment is assumed to be constant at different temperatures, which results in the same fluence at the target. Therefore, the $PA_{signal}$ becomes constant according to Eq. 2. We also observed a decrease in the background noise ($PA_{noise}$) when the temperature droped, and that resulted in a SNR increase.

Fig. 2(b) shows the results for the second set of simulations where the scattering coefficient decreases with lowering the temperature with the rate 0.4% $°C^{-1}$ [24] while Grüneisen parameter remains constant. However these results demonstrate SNR improvement of PA signal with lowering the temperature, the improvement is less than that obtained in the first set of simulations. We observed that $PA_{signal}$ increases at lower temperatures due to the greater fluence (as a results of lower light scattering) reaching the target. On the other hand, the background noise increases with lowering the temperature. Both effects become more significant at greater depths.

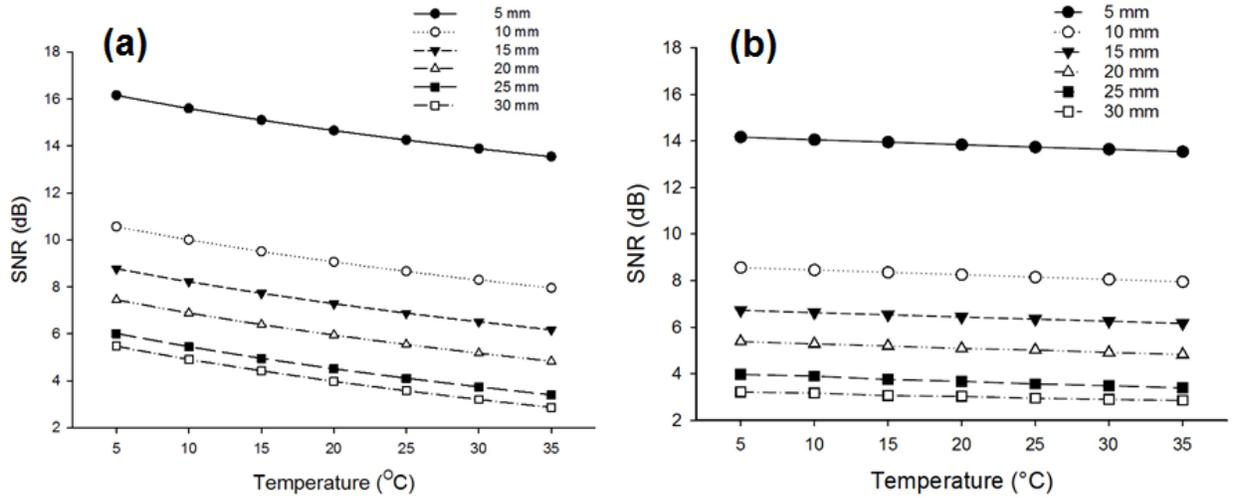

**Figure 2.** Simulation results of SNR improvement of PA signal with lowering the temperature at different temperatures and depths. (a) Scattering coefficient is held constant while Grüneisen parameter is changed with temperature. (b) Scattering coefficient is changed with temperature while Grüneisen parameter is held constant.

In the third set of simulations, the effect of temperature on Grüneisen parameter and scattering coefficient are taken into account (Fig. 3). These simulations demonstrate a better improvement in the SNR compared to the results in Fig. 2(a) and (b).

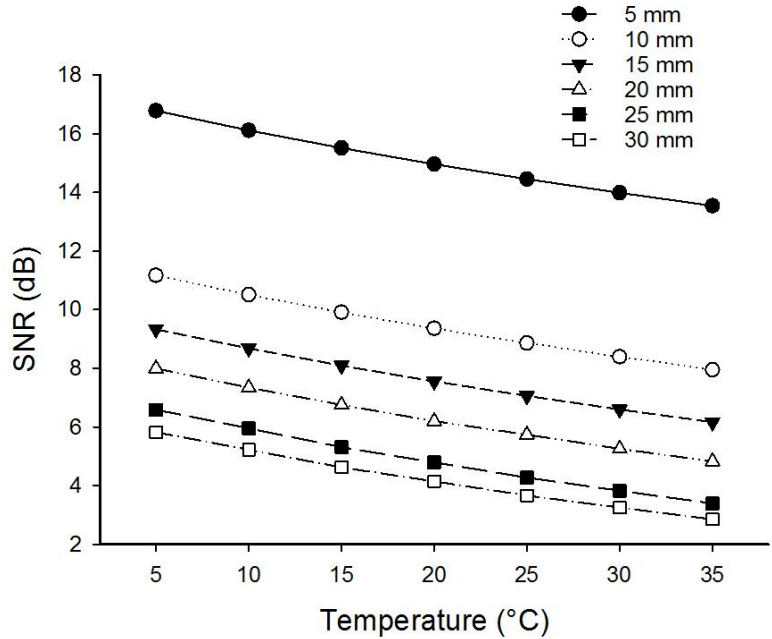

**Figure 3.** Simulation results of SNR improvement of PA signal with lowering the temperature at different temperatures and depths. Both scattering coefficient and Grüneisen parameter are changed with temperature.

**Experimental results**

We first explored the temperature dependence of photoacoustic imaging in a water phantom. The experimental setup in this case is the same as shown in Fig. 1 without chicken breast tissue. With this initial phantom, we then examined the depth-temperature dependence of PA signal SNR within a chicken breast sample. This study is to determine the practicality of the proposed method in biological tissues. Wayne State University Animal Care and Use Committee has approved the study. All methods are performed in accordance with the relevant guidelines and regulations. In the current work, the temperature of the imaging target i.e., covered resistance heating wire, is kept constant by thermal control of the target through an external circuit at 37°C. The temperature of the intermediate medium, on the other hand, is decreased gradually by implementing the experiment in a cooling water bath.

Figs. 4 (a) and (b) present the PA images at 10°C and 35°C (of the intermediate media) projected onto the ultrasound images, when the imaging target is placed 30 mm inside the water and chicken tissue, respectively. During these measurements the temperature of the imaging target (the coated black wire) is kept constant at 37°C by thermal control of the target medium through an external circuit. In both cases, the low-temperature mediated enhancement of the photoacoustic imaging can be easily observed by comparing the PA images at 10°C and 35°C. We used a double stage delay multiply and sum beamforming algorithm [45, 56] for image reconstruction.

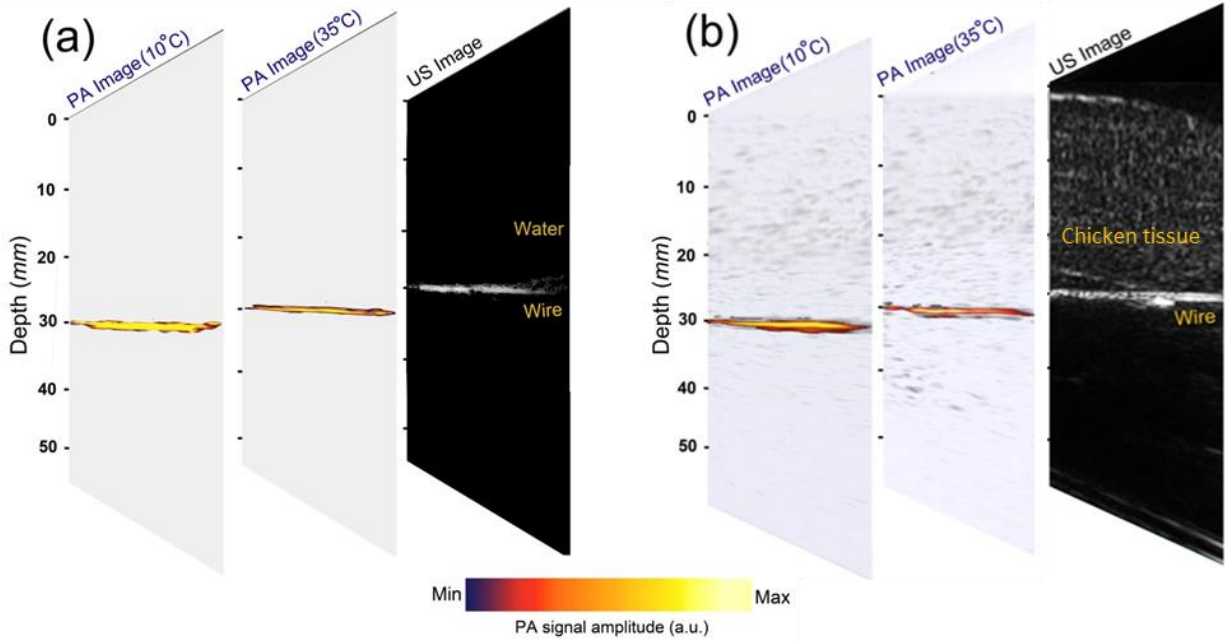

**Figure 4.** The PA images projected on the ultrasound (US) image is presented for the case of 30 mm imaging depth at low (10°C) and high (35°C) temperatures. The imaging target temperature during the imaging process is kept constant at 37°C. The intermediate medium is water in **(a)**, and chicken breast tissue in **(b)**.

To be able to provide a quantitative comparison between the obtained images, we report, in Fig. 5, the SNR from the individual photoacoustic images. This value is calculated as:

$$SNR = 10 Log_{10} \left( \max(I^2) / \sigma_b^2 \right) \qquad (11)$$

where $I$ and $\sigma_b^2$ indicate the peak intensity and the variance of the background in the image, respectively. The paramount feature in Fig. 5 is the enhancement of the SNR when the temperature of the intermediate medium is decreased. The SNR value is calculated for 100 frames in each experiment.

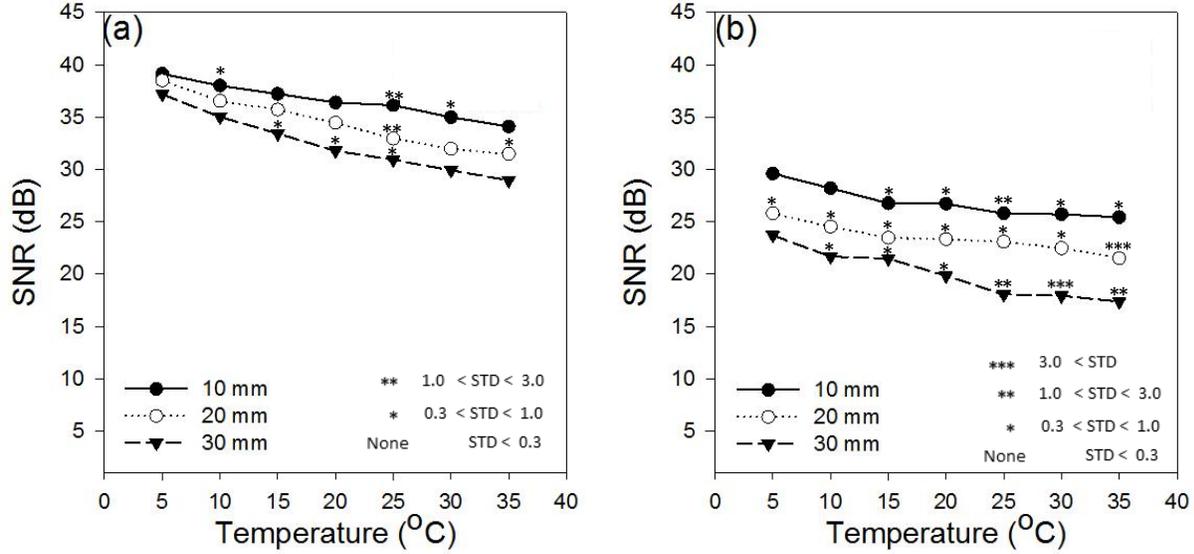

**Figure 5.** SNR of the imaging target (resistance heating wire) versus temperature at three different depths (10mm, 20mm, and 30mm) inside **(a)** water, and **(b)** chicken breast tissue. STD stands for standard deviation.

According to the simulation results (Fig. 3), the changes of Grüneisen parameter with temperature is the dominant factor. In the first experiment, the intermediate medium is water, and in the second experiment is chicken breast tissue. The Grüneisen parameter of water decreases more rapidly compared to that of the tissue. That is probably due to the different tissue compositions' thermal expansion coefficients in chicken breast tissue which are sometimes opposite [17]. This means that the background noise is also decreased more rapidly with lowering the temperature in water compared to that in the tissue, hence a more rapid change of SNR in the water experiment is seen.

Moreover, the absorption coefficient of water is varying linearly with temperature and becoming smaller as temperature decreases [40, 41]. When the intermediate medium is only water (Figure 5(a)), this change in absorption, although small, can affect the SNR in two ways: first, by increasing the PA signal generated from the medium which contributes into noise, and second by increasing the light attenuation and decreasing the PA signal generated from the target. In the tissue, water is not the main absorber and its absorption coefficient change is insignificant compared to the total attenuation coefficient of the tissue. This might be another reason for having smaller SNR changes in tissue compared to water.

There are other methods for deep PA imaging such as optical clearing of bio-tissues. In this method, the scattering coefficient and the degree of forward scattering of photons are manipulated, by administering some chemicals to increase the light penetration inside the tissue [46-48]. This method is invasive and cannot easily be translated to the clinic [46]. Other methods based on wavefront engineering are computationally expensive and require costly setup [50]. In several studies, contrast agents have been utilized to increase the penetration depth of the PA imaging system [51,52]. Such methodologies make PA imaging an invasive technique. The method we explained here is simple and does not require any special equipment.

Since the suppressed laser absorption through the intermediate medium is a fundamental molecular characteristic, any photoacoustic configuration including different types of photoacoustic microscopy (PAM) and photoacoustic tomography (PAT) [53] can benefit from it.

The proposed approach provides a benchmark demonstration on improving the signal strength due to the two-step excitation of the absorber medium on one hand, and a better imaging depth due to the cooling of the intermediate medium, on the other hand. Laser cooling has not yet been extended to biological molecules because of their complex structures. However, this complexity makes biomolecules potentially good candidates for this purpose, due to possessing permanent electric dipole moments that lead to long-range, tunable, anisotropic dipole–dipole interactions [54].

Some of the future studies include, developing PA spectroscopy to find the optimum wavelength at which the proposed method is most efficient; implementing laser refrigeration setup and creating a low-temperature channel rather than bringing down the temperature of the whole intermediate media; exploring the most optimum ultrasound frequency in the detection unit; and exploring laser refrigeration in different configurations of PAI, i.e., PAM and PAT.

**Conclusion**

We demonstrated, through several sets of simulation and experiment, that the SNR of the photoacoustic signal generated from a black wire as an imaging target at different depths, is improved by lowering the temperature of the intermediate medium (from 35°C to 5°C) between the PA sensor and the target. Initially water was used as the intermediate medium, then chicken breast tissue. Simulations showed that the improvement is mainly due to the reduction of Grüneisen parameter of the medium which leads to a lower level of background noise. They also showed a decrease in the scattering coefficient of the intermediate medium which further improved the SNR of the PA signal. Since there was no evidence about how the absorption coefficient of tissue changes with temperature, particularly in the range of temperatures investigated in this study, the temperature dependence of the PA signal strength due to changes in absorption coefficient was not investigated.

Compared to other depth improvement methods in PA imaging, the method presented here does not require a complicated setup and it is still noninvasive. These findings may open new possibilities toward the application of biomedical laser refrigeration.

**Acknowledgement**

The authors acknowledge Wayne State University Startup fund.

**Author contributions**

M.R.N.A. and H.Z.J. conceived the study. H.Z.J., A.H., Q.X., and Y.Z. conducted the experiments. S.M., H.Z.J., Q.X., M.A.A. and M.R.N.A. analyzed the data. H.Z.J., S.M., A.H., Q.X., and M.R.N.A. wrote the manuscript and all authors participated in paper revisions.

**Competing financial interests:**

The authors declare no competing interests.